\documentclass[aps,floatfix,twocolumn,pra,superscriptaddress]{revtex4}

\usepackage{epsfig,amsmath,graphicx,graphics,float}
\usepackage{subfigure}
\usepackage{color}
\usepackage{latexsym}

\begin{document}

\def\simge{\mathrel{%
         \rlap{\raise 0.511ex \hbox{$>$}}{\lower 0.511ex \hbox{$\sim$}}}}
\def\simle{\mathrel{
         \rlap{\raise 0.511ex \hbox{$<$}}{\lower 0.511ex \hbox{$\sim$}}}}
\newcommand{\feynslash}[1]{{#1\kern-.5em /}}
\newcommand{\iac}[1]{{\color{red} #1}}

\title{Chandrasekhar-Clogston limit and critical polarization in a Fermi-Bose superfluid mixture}

\author{Tomoki Ozawa}
\affiliation{
INO-CNR BEC Center and Dipartimento di Fisica, Universit\`a di Trento, I-38123 Povo, Italy
}%

\author{Alessio Recati}
\affiliation{
INO-CNR BEC Center and Dipartimento di Fisica, Universit\`a di Trento, I-38123 Povo, Italy
}%

\author{Marion Delehaye}
\affiliation{
Laboratoire Kastler-Brossel, \'Ecole Normale Sup\'erieure, CNRS and UPMC, 24 rue Lhomond, 75005 Paris, France
}

\author{Fr\'ed\'eric Chevy}
\affiliation{
Laboratoire Kastler-Brossel, \'Ecole Normale Sup\'erieure, CNRS and UPMC, 24 rue Lhomond, 75005 Paris, France
}

\author{Sandro Stringari}
\affiliation{
INO-CNR BEC Center and Dipartimento di Fisica, Universit\`a di Trento, I-38123 Povo, Italy
}%

\date{\today}

\begin{abstract}
We study mixtures of a population-imbalanced strongly-interacting Fermi gas and of a Bose-Einstein condensed  gas at zero temperature. In the homogeneous case, we find that the Chandrasekhar-Clogston critical polarization for the onset of instability of  Fermi superfluidity  is enhanced due to the interaction with the bosons. Predictions for the critical polarization are also given in the trapped case, with a special focus to the situation of equal Fermi-Bose and Bose-Bose coupling constants, where the density of fermions becomes flat in the center of the trap. This regime can be realized experimentally using Feshbach resonances and  is well suited to investigate the emergence of exotic configurations, such as the occurrence of spin domains or the FFLO phase.
\end{abstract}

\maketitle

\section{Introduction}

The property of fermions interacting  with a Bose fluid has been a long-standing subject of research in condensed matter physics, dating back to the study of $^3$He-$^4$He mixtures~\cite{Bardeen1966}.
With the recent development of research activity in ultracold gases, it is now possible to experimentally create mixtures of degenerate bosonic and fermionic atomic gases~\cite{Schreck2001, Hadzibabic2002, Inouye2004, Gunter2006, Ospelkaus2006, Zaccanti2006, Best2009, Sugawa2011, Wu2011, Schreck2013}.
Very recently, the first experimental realization of a superfluid Bose-Fermi mixture was reported~\cite{FerrierBarbut2014}, the Fermi gas being at the unitarity limit.
 
There are several theoretical works on mixtures of superfluid Bose gases interating with spin-1/2 Fermi gases~\cite{Modak2011, Yamamoto2012, Bukov2014, Maeda2009, Ramachandhran2011}, but the behavior of coexisting superfluid Fermi and Bose gases in the case of strong Fermi-Fermi interaction has not yet been considered in the literature.
Furthermore, since spin-imbalanced fermions are predicted to give rise to exotic phases such as the Fulde-Ferrell-Larkin-Ovchinnikov (FFLO) phase~\cite{Hu2006, Bulgac2008,Radzihovsky2010}, it is of great interest to investigate how their behavior is modified by the interaction with bosons. 

In this paper, we show that in a homogeneous configuration the Chandrasekhar-Clogston critical polarization for the breakdown of superfluidity  is larger than in the absence of the bosonic component ~\cite{Recati2008}.
We then consider the case of  a  harmonically trapped configuration: when the Bose-Bose and  Bose-Fermi interactions are equal, the fermionic density in the region of coexistence  with bosons becomes flat, because the interaction with bosons exactly compensates the external trapping potential~\cite{Molmer1998}.
We investigate the phase diagram of the trapped gas when the fermion imbalance is varied, and show that, for a finite range of polarization, the fermionic density in the Bose-Fermi coexistence region can become inhomogeneous.

\section{Homogeneous system}
\begin{figure}[htbp]
\begin{center}
\includegraphics[width=8.0cm]{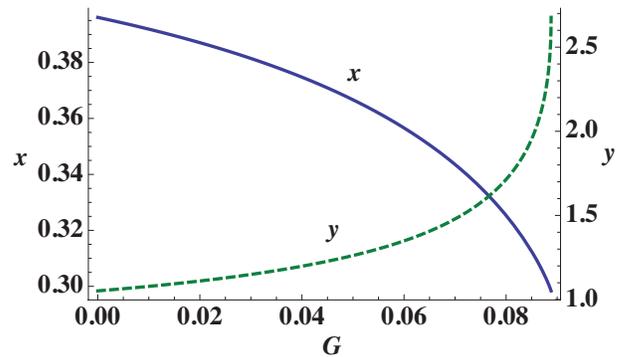}
\caption{Critical ratios $x \equiv n_\downarrow/n_\uparrow$ (solid blue line with left axis) and $y \equiv n_s/n_\uparrow$ (dotted green line with right axis) as a function of $G \equiv n_\uparrow g_{bf}^2/\epsilon_{F\uparrow} g_{bb}$.}
\label{xandy}
\end{center}
\end{figure}

The balanced unitary Fermi gas is known to be fully superfluid at zero temperature.
As one increases the polarization, it has been observed that the system phase separates into a balanced superfluid phase and an imbalanced normal phase~\cite{Shin2008}.
The two phases have different densities, and the equilibrium conditions between the two phases fix the ratio $x$ between the density of the minority species over the density of the majority species in the normal phase, which determine the Chandrasekhar-Clogston limit. At zero temperature, this critical ratio turns out to be, at unitarity, $x\approx 0.4$~\cite{Lobo2006, Recati2008, ZwergerBook}.
As we show, this value is modified by the interaction with bosons. We assume that the Fermi gas is phase separated into a superfluid phase with density $n_s$ for both species and a normal phase with density $n_\uparrow$ and $n_\downarrow$ for the spin-up (majority) and spin-down (minority) fermions, respectively.
The density of the coexisting bosons in the Fermi superfluid phase is $n_{bs}$ and that in the normal phase is $n_{bn}$. Later we discuss the stability conditions for such configurations.
We assume that both the bosonic and fermionic species can be described within the local density approximation and both the Bose-Bose and the Bose-Fermi interactions are weak enough to be treated within the mean field approximation.
Then the energy density in the superfluid phase ($E_s$) and in the normal phase ($E_n$) takes the form
\begin{align}
	E_s &= \frac{g_{bb}}{2}n_{bs}^2 + 2 g_{bf} n_{bs} n_s + e_s [n_s], \notag \\
	E_n &= \frac{g_{bb}}{2}n_{bn}^2 + g_{bf} n_{bn} (n_\uparrow + n_\downarrow) + e_n [n_\uparrow, n_\downarrow],
\end{align}
where $g_{bb} \equiv 4\pi \hbar^2 a_{bb}/m_b$, assumed to be positive, and $g_{bf} \equiv 2\pi \hbar^2 a_{bf}/m_r$ are, respectively, the Bose-Bose and spin-independent Bose-Fermi interaction coupling constants.
The Bose-Bose and Bose-Fermi scattering lengths are $a_{bb}$ and $a_{bf}$, respectively, and $m_r \equiv m_b m_f/(m_b + m_f)$ where $m_b$ and $m_f$ are the boson and fermion masses, respectively. 
The Fermi energy density in the superfluid phase is given by the universal form
\begin{equation}
	e_s [n_s] \equiv \xi \frac{6}{5}\frac{\hbar^2}{2m_f}(6\pi^2 n_s)^{2/3}n_s,
\end{equation}
where $\xi = 0.370$~\cite{ZwergerBook, Ku2012, Zurn2013} is the Bertsch parameter.
For the normal phase we use the  expansion in the parameter $x \equiv n_\downarrow / n_\uparrow$ introduced in ~\cite{Recati2008}:
\begin{align}
		\notag \\
	e_n [n_\uparrow, n_\downarrow] &\equiv \frac{3}{5} \epsilon_{F\uparrow} n_\uparrow \left( 1 - \frac{5}{3}Ax + \frac{m_f}{m^*} x^{5/3} + F x^2 \right)
	\notag \\
	&\equiv \frac{3}{5} \epsilon_{F\uparrow} n_\uparrow \epsilon (x),
\end{align}
 where $\epsilon_{F\uparrow} \equiv (\hbar^2 / 2m_f) (6\pi^2 n_\uparrow)^{2/3}$ is the non-interacting Fermi energy of the majority species and for the parameters in $\epsilon (x)$ we use $A = 0.615$, $m^*/m_f = 1.20$, and $F = (5/9)A^2$, determined by diagramatic methods and Monte-Carlo calculations~\cite{Prokof'ev2008, Combescot2008, Mora2010}. Using different sets of parameters would not change our results significantly. The equilibrium between the two phases is determined by matching the pressure and the chemical potentials for both bosons and fermions at the interface, which leads to the following conditions for $x$ and $y \equiv n_s/n_\uparrow$:
\begin{align}
	&\xi y^{2/3} - 2G y
	-
	\frac{1}{2}\epsilon(x) - \frac{3}{10}\epsilon^\prime (x) (1-x) + G (1+x)=0,
	\notag \\
	&2G y^2 - \frac{4}{5}\xi y^{5/3}
	-
	G \frac{(1+x)^2}{2} + \frac{2}{5}\epsilon (x)=0
	,\label{xyconditions}
\end{align}
where $\epsilon^\prime (x) \equiv d\epsilon (x)/dx$ and $G \equiv n_\uparrow g_{bf}^2/(\epsilon_{F\uparrow} g_{bb})$ is a dimensionless parameter independent of the bosonic density. As a consequence also the critical ratios $x$ and $y$ are independent of the boson density, provided there are background bosons with nonzero densities in both phases. The parameter $G$ has an important physical meaning, corresponding to the ratio between the change in the energy of fermions caused by the induced interaction $-g_{bf}^2 /g_{bb}$ in the static limit and the non-interacting Fermi energy.
The existence of two real solutions for $x$ and $y$ for (\ref{xyconditions}) is ensured for $0 \le G \le G_{\mathrm{max}} \approx 0.089$ and in Fig.~\ref{xandy} we plot the resulting values  of  $x$ and $y$ as a function of $G$.
When $G = 0$, the critical ratio $x \approx 0.40$ coincides with the value obtained in the absence of Bose-Fermi interaction ($g_{bf}=0$).
As $G$ becomes larger, the value of $x$ decreases reaching the minimum value of $x \approx 0.30$, which means that the superfluid phase of fermions is stabilized by the interaction with bosons.
The ratio $y$, on the other hand, increases with $G$ reaching the maximum value of $y \approx 2.68$, which implies that the density jump at the interface of the two phases becomes larger; the maximum value of the jump, corresponding to $G= G_{\mathrm{max}}$, is  $2 n_s / (n_\uparrow + n_\downarrow) \approx 4.1$ to be compared with the value $\approx 1.5$ when $G=0$.

The nonexistence of real solutions when $G > G_{\mathrm{max}}$ is related to the occurrence of  dynamical instability in the fermionic superfluid phase caused by the interaction with bosons.
The dynamical stability of the superfluid phase requires that the following inequality be obeyed~\cite{Viverit2000}:
\begin{align}
	\frac{\delta^2 e_s [n_s]}{\delta n_s^2} - 4 \frac{g_{bf}^2}{g_{bb}} > 0,
\end{align}
which is equivalent to imposing $\xi / 3 y^{1/3} > G$. 
We have checked that the condition for having  real solutions  for $x$ and $y$  coincides with the one ensuring  dynamical stability. 
If $G$ becomes larger than $G_{\mathrm{max}}$, the superfluid Fermi gas and the Bose gas are expected to phase separate.

\section{Trapped system}
Let us  now consider the case of a trapped quantum mixture. In the absence of bosons, it is known that as one introduces a small imbalance between the two species, the central part of the trap remains superfluid, and the outer shell is turned into a normal state~\cite{Shin2008,Recati2008}. When the imbalance is large enough, the whole Fermi gas is in the normal state.

In the presence of bosons, the situation can change significantly.
The energy of a highly polarized Fermi gas interacting with a BEC gas is given, within the local density approximation (LDA), by
\begin{align}
	&E
	=
	\int_{ r < R_b}d^3 r
	\left\{
	\frac{g_{bb}}{2}n_b^2 (r) + [V_b (r) - \mu_b] n_b (r)
	\right.
	\notag \\
	&\hspace{1cm}\left.
	+ g_{bf} n_b (r) (n_\uparrow (r) + n_\downarrow (r)) + e_n [n_\uparrow (r), n_\downarrow (r)]
	\right.
	\notag \\
	&\hspace{1cm}
	\left.
	+ [V_f (r) - \mu_\uparrow ]n_\uparrow (r) + [ V_f (r) - \mu_\downarrow ]n_\downarrow (r)
	\right\}
	\notag \\
	&+
	\int_{R_b < r}d^3 r
	\left\{
	e_n [n_\uparrow (r), n_\downarrow (r)]
	+ [V_f (r) - \mu_\uparrow ]n_\uparrow (r)
	\right.
	\notag \\
	&\hspace{4cm}
	\left.+ [ V_f (r) - \mu_\downarrow ] n_\downarrow (r)
	\right\}, \label{enormal}
\end{align}
where $R_b$ is the radius at which the boson density vanishes and $V_b(r)$ and $V_f(r)$ are the harmonic traps for bosons and fermions, respectively~\footnote{In the following, without loss of generality, we use an isotropic potential}.
The densities of boson, spin-up fermion, and spin-down fermion are $n_b (r)$, $n_\uparrow (r)$, and $n_\downarrow (r)$, respectively, and the corresponding chemical potentials are labelled, respectively, with  $\mu_b$, $\mu_\uparrow$, and $\mu_\downarrow$.

Taking the variation of the energy with respect to $n_b (r)$, $n_\uparrow (r)$, and $n_\downarrow(r)$ in the Bose-Fermi coexistence region $r < R_b$, which will be hereafter referred to as the ``core" region, one obtains the following equations:
\begin{align}
	&n_b (r) = \left[ \mu_b - V_b(r) - g_{bf} (n_\uparrow (r) + n_\downarrow (r))\right]/g_{bb},
	\notag \\
	&\frac{\delta e_n}{\delta n_\sigma} + V_f (r) - \frac{g_{bf}}{g_{bb}} V_b (r) + \frac{g_{bf}}{g_{bb}}\mu_b - \mu_\sigma 
	\notag \\
	&\hspace{2cm}- (g_{bf}^2/g_{bb})(n_\uparrow (r) + n_\downarrow (r))
	= 0, \label{eqcore}
\end{align}
where $\sigma = \uparrow,\downarrow$. The second equation explicitly reveals that, if $g_{bf}V_b(r) = g_{bb}V_f(r)$, the fermion densities are not affected by the presence of the trap, and take a constant value inside the core~\cite{Molmer1998}. This follows from the fact that the effect of the  trap on the fermions is exactly canceled by the mean-field interaction with bosons~\footnote{The cancellation effect disappears when the number of bosons is not large enough to apply LDA.}.  Conversely the bosonic density is not affected by the presence of fermions and, choosing an external potential of harmonic form, the bosonic density, for $r < R_b$, takes an inverted parabola profile, whose shape is solely determined by the total number of bosons and the Bose-Bose coupling constant $g_{bb}$. If instead $g_{bf} V_b > g_{bb} V_f$, the fermions feel an anti-trapping potential in the core region and their density will increases when one moves away from the center.

When the imbalance is small, most of the fermions are in the superfluid phase, and
one can write down a similar energy functional as (\ref{enormal}), but the region $r < R_b$ is filled with the superfluid phase, while the region $r > R_b$ is divided into an inner superfluid phase and an outer normal phase.
One obtains the following conditions analogous to Eq.~(\ref{eqcore}):
\begin{align}
	&n_b (r) = \left[ \mu_b - V_b(r) - 2 g_{bf} n_s (r)\right]/g_{bb},
	\notag \\
	&\frac{\delta e_s}{\delta n_s} + 2 \left( V_f(r) - \frac{g_{bf}}{g_{bb}} V_b(r)\right) + 2\frac{g_{bf}}{g_{bb}}\mu_b - (\mu_\uparrow + \mu_\downarrow)
	\notag \\
	&\hspace{4cm}- 4(g_{bf}^2/g_{bb})n_s (r)
	= 0, \label{eqcoresf}
\end{align}
in the core.
As in the highly polarized case, one can see that in this region the fermions exhibit a flat density distribution when $g_{bf}V_b(r) = g_{bb} V_f (r)$.
The equilibrium between the superfluid phase and the normal phase in the tail is determined by matching the pressure and the chemical potentials at the interface, and the critical ratio $x = n_\downarrow/n_\uparrow$ is equal to 0.40, which is the value predicted in the absence of bosons~\cite{Lobo2006, Recati2008, ZwergerBook}.

For concreteness we provide predictions for the mixture of $^7$Li bosons and $^6$Li fermions reported in~\cite{FerrierBarbut2014} where $V(r) \equiv V_b(r) = V_f (r)$, and we focus on the special case  $g_{bf} = g_{bb}$. This condition $g_{bb}=g_{bf}$ (corresponding to $ a_{bb}/a_{bf}= (m_b+m_f)/2m_f$), together with that of unitarity for the Fermi component, are achievable for a magnetic field of $B = 817$ G, leading to a fermion-fermion scattering length of $25 800a_B$, and to a boson-boson scattering length $a_{bb}= 44.2 a_B$, the average Fermi momentum being $k_F = 10^6 \sim 10^7$ m$^{-1}$. 

\begin{figure}[htbp]
\begin{center}
\subfigure[The whole system is normal]{
\includegraphics[width=4.1cm]{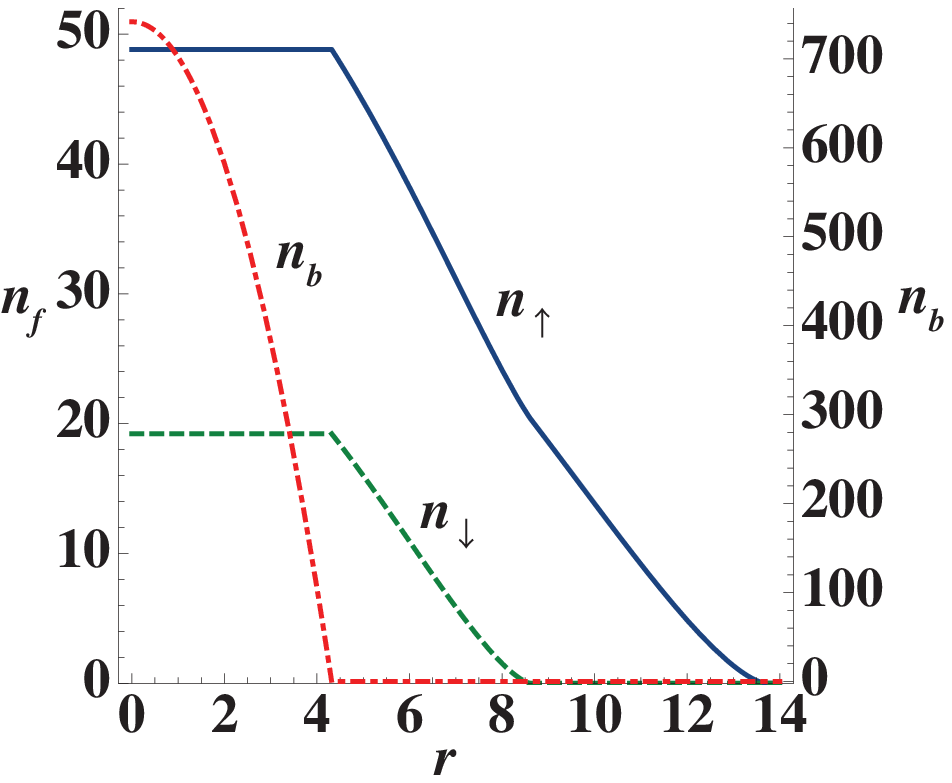}}
\subfigure[The core is all superfluid]{
\includegraphics[width=4.1cm]{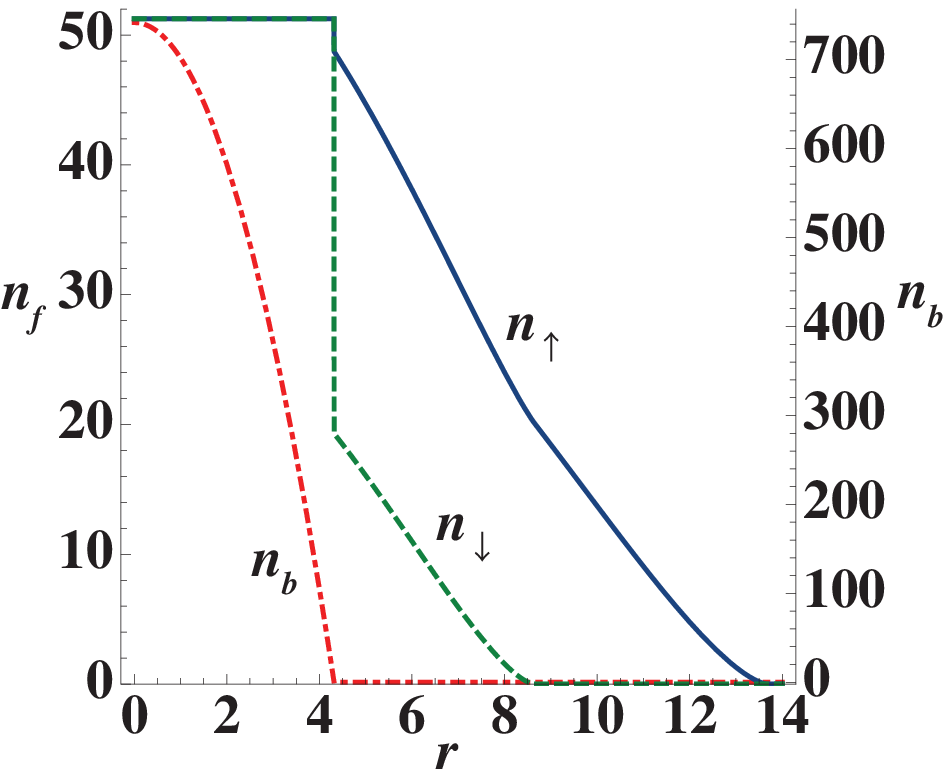}}
\caption{Local 3D density profile of the two opposite limits where the inhomogeneous phase in the core is about to appear. We fix $N_b = 10^5$ and $N_\uparrow = 1.5 \times 10^5$. The solid (blue) lines are spin-up fermions, the dotted (green) lines are spin-down fermions, and the dash-dotted (red) lines are bosons.
The left axis is for the fermion densities and the right axis is for the boson density.
The number of spin-down fermions is (a) $N_\downarrow = 0.22\times10^5$; (b) $N_\downarrow = 0.33\times 10^5$. The length is in units of $l_{\mathrm{ho}}$, and the density of particles is in units of $1/l_{\mathrm{ho}}^3$.}
\label{verge}
\end{center}
\end{figure}

\begin{figure}[htbp]
\begin{center}
\includegraphics[width=8.0cm]{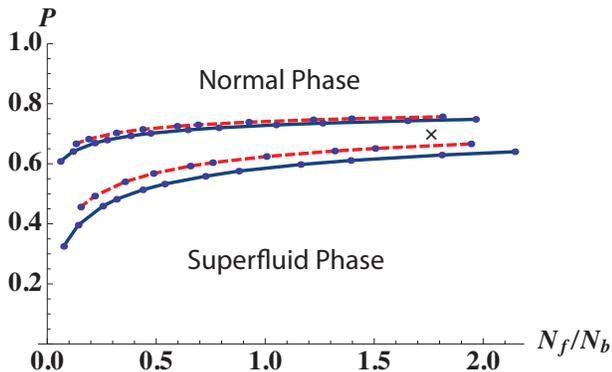}
\caption{Critical polarizations for entering the inhomogeneous core as a function of $N_f/N_b$ for $N_b = 10^5$ (solid blue lines) and $N_b = 10^4$ (dashed red lines). The cross corresponds to the situation of Fig.~\ref{bosons_fermions_snn}.}
\label{polarization}
\end{center}
\end{figure}

The density profile of the fermions (both inside and outside the core) can be obtained by solving (\ref{eqcore}) or (\ref{eqcoresf}) and similar equations for the region $r > R_b$.
In Fig.~\ref{verge}, we plot two density distributions for fixed values of $N_b = 10^5$ and $N_\uparrow = 1.5 \times 10^5$ but with two different values of $N_\downarrow$.
We choose $a_{bb} = 10^{-3} l_{\mathrm{ho}} m_b/m_f$, where $l_\mathrm{ho} \equiv \sqrt{\hbar / m_f \omega_f}$ is the harmonic oscillator length corresponding to a fermionic trap frequency $\omega_f =2\pi \cdot 420$Hz. Fig.~\ref{verge}(a) corresponds to the smallest value of total polarization of the gas ($P \equiv(N_\uparrow-N_\downarrow)/(N_\uparrow+N_\downarrow) = 0.74$) compatible with the absence of superfluidity, where the ratio $n_\downarrow/n_\uparrow$ in the core is equal to the critical value determined by Eq.~(\ref{xyconditions}) for the value of $G$ in the core region. A smaller value of $P$ would correspond to the onset of a superfluid region in the core. Fig.~\ref{verge}(b) instead corresponds to the largest value of total polarization ($P=0.63$) compatible with the presence of a superfluid phase occupying the whole core region. A larger value  of $P$ would correspond to the onset of a normal region in the core (see also Fig. \ref{polarization}). 

For intermediate values of the population imbalance, coexistence of the superfluid and the normal phase takes place in the core region, giving rise to inhomogeneity and new interesting physics.
Inhomogeneity in the core can be reached either by starting with a balanced superfluid gas and gradually decrease the number of minority fermions till the normal part enters the core, or by starting with a completely polarized gas and gradually increase the number of minority fermions till a superfluid phase region in the core is favorable. In Fig.~\ref{polarization}, the two critical polarizations for entering the inhomogeneous core phase are plotted as a function of $N_f/N_b$, where $N_f \equiv N_{\uparrow}+N_{\downarrow}$, for two different values of $N_b$. The upper region corresponds to the phase with the whole system being normal (Fig.~\ref{verge}(a)), and the lower region corresponds to the whole core being superfluid (Fig.~\ref{verge}(b)).
The region between the lines represents the inhomogeneous core phase.
We observe that the critical polarization as a function of $N_f/N_b$ is not very sensitive to the number of bosons.
The two critical polarization lines approach the value $0.8$ as $N_f/N_b \to \infty$.
This asymptotic value corresponds to the critical polarization for the onset of superfluidity in the absence of bosons~\cite{Lobo2006}.

\begin{figure}[htbp]
\begin{center}
\subfigure[Local (left) and doubly integrated (right) density for the S-N scenario (see text)]{
\includegraphics[width=8.5cm]{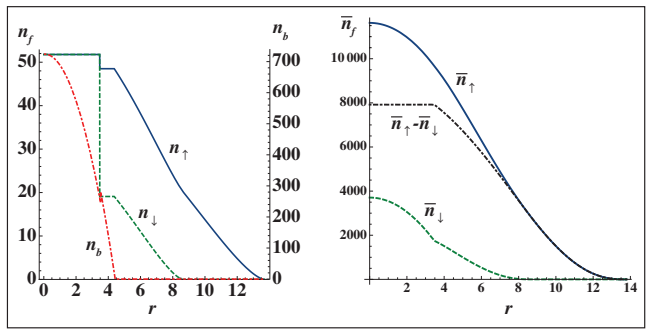}}
\subfigure[Local (left) and doubly integrated (right) density for the N-S-N scenario (see text)]{
\includegraphics[width=8.5cm]{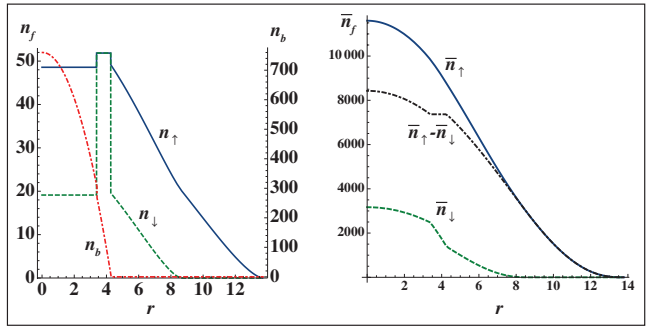}}
\caption{Local 3D density and doubly integrated density profiles for two different configurations for the core, corresponding to the S-N and N-S-N scenarios in the text.
We have chosen $N_b = 10^5$ and $N_\uparrow = 1.5 \times 10^5$ as in Fig.~\ref{verge}.
The value of $N_\downarrow$ is instead $0.28\times 10^5$, corresponding to $P = 0.69$ and $N_f/N_b = 1.78$, i.e. to the inhomogeneous core region of Fig~\ref{polarization}.
The solid (blue) lines are for spin-up fermions and the dotted (green) lines are for spin-down fermions. The dash-dotted (red) lines are for bosons for the local density, and the dash-dotted (black) lines are the difference $\bar{n}_\uparrow - \bar{n}_\downarrow$ for the doubly integrated density. For the local density, the left axis is for fermions and the right axis is for bosons. Lengths are in units of $l_{\mathrm{ho}}$.}
\label{bosons_fermions_snn}
\end{center}
\end{figure}

We now discuss the possible scenarios characterizing the inhomogeneous phase for intermediate values of population imbalance (see Fig.~\ref{polarization}). The simplest possibility, hereafter called the superfluid-normal (S-N) scenario, is that the core is phase separated into a central superfluid and an outer normal phase. The equilibrium condition between the superfluid phase and the normal phase turns out to be determined by the same conditions (\ref{xyconditions}) holding for the homogeneous mixture.
Another possibility, hereafter called the normal-superfluid-normal (N-S-N) scenario, is that the core is phase separated into a central normal phase and an outer superfluid phase, while the tail is normal.
The two scenarios have very similar energies, and can be easily distinguished in experiments~\footnote{An experimental measurement of the local 3D density would require a more sophisticated 3D reconstruction of the density based on inverse Abel transformation~\cite{Shin2008}.} by measuring the doubly-integrated column density $\bar{n}_\sigma (z) \equiv \int dx dy\, n_\sigma(x,y,z)$, because the superfluid region appears as a flat profile in the difference $\bar{n}_\uparrow (z) - \bar{n}_\downarrow (z)$~\cite{DeSilva2006PRA}. This flat doubly-integrated density profile is due to pairing and should not be confused with the 3D flat density profile that is caused by the Fermi-Bose interaction.
Typical density distributions and corresponding doubly integrated column densities are plotted in Fig.~\ref{bosons_fermions_snn}. Another interesting feature of this inhomogeneous core phase is that the boson density is not a simple inverse parabola, but has a small jump (not visible in the figure) at the phase boundary between the superfluid and normal fermion.
The two scenarios of Fig.~\ref{bosons_fermions_snn} can be energetically separated by changing the value of $g_{bb}$ as compared to $g_{bf}$. If, e.g., $g_{bb} \lesssim g_{bf}$ the fermions feel a small anti-trapping potential in the core, and the second scenario, Fig.~\ref{bosons_fermions_snn}(b), will take place. The difference should be clearly visible experimentally, as shown in the doubly integrated densities in Fig.~\ref{bosons_fermions_snn}.

\par The emergence of  the inhomogeneous phase is also compatible with other more exotic possibilities, like the emergence of the FFLO phase~\cite{Hu2006, Bulgac2008, Radzihovsky2010}. Indeed the local chemical potential for fermions is constant over the flat region, therefore phases which can exist only within a narrow range in the chemical potential could be observed in the core.

\begin{acknowledgments}
We thank Igor Ferrier-Barbut, and Christophe Salomon for useful discussions.
We also thank Stefano Giorgini for insightful comments.
This work was supported by the ERC through the QGBE grant and by Provincia Autonoma di Trento, and by the ERC Ferlodim and Thermodynamix, the Ile de France Nano-K (contract Atomix) and Institut de France Louis D. Prize.
\end{acknowledgments}

\end{document}